# Parallels and promising directions in the study of genetic, cultural, and moral evolution

**Rohan Maddamsetti**

Department of Biological Sciences, Old Dominion University, Norfolk, VA

**Jacob Bower-Bir**

Ostrom Workshop in Political Theory and Policy Analysis, Indiana University, Bloomington, IN

**Abstract**

Experimental evolution has yielded surprising insights into human history and evolution by shedding light on the roles of chance and contingency in history and evolution, and on the deep evolutionary roots of cooperation, conflict and kin discrimination. We argue that an interesting research direction would be to develop computational and experimental systems for studying evolutionary processes that involve multiple layers of inheritance (such as genes, epigenetic inheritance, language, and culture) and feedbacks (such as gene-culture coevolution and mate choice) as well as open-ended niche construction—all of which are important in human history and evolution. Such systems would also be a clear way to motivate evolution and computation to scholars and students across diverse cultural and socioeconomic backgrounds, as well as to scholars and students in the social sciences and humanities. In principle, computational models of cultural evolution could be compared to data, given that large-scale datasets already exist for tracking cultural change in real-time. Thus, experimental evolution, as a laboratory and computational science, is poised to grow as an educational tool for people to question and study where we come from, why we believe what we believe, and where we as a species may be headed.

## Human history in light of experimental evolution

Can evolution experiments help at all in the interpretation of human population genomics and history? Surprisingly, in some cases the answer may be yes.

Reports that a small number of powerful men, possibly including Genghis Khan, have fathered millions of male descendants have garnered much attention in the popular press [1-3]. It seems that warlords with astounding reproductive success have left their footprint on the evolutionary record as well as history [4]. What is truly surprising however, is that similar evolutionary dynamics occur during the exponential growth of spatially structured populations of cells, such as in biofilms, tumors, and developing embryos—in the absence of natural selection. The key mathematical connection is that the distribution of offspring among people is heavy-tailed, such that some people are born into notably large families. The same goes for lineages of cells that are growing exponentially in spatially-structured populations.

## *Heavy-tailed offspring distributions in expanding populations can cause fictitious selection pressures*

In a series of elegant theoretical and experimental investigations, Oscar Hallatschek and colleagues have shown that in expanding bacterial colonies founded by a clonal ancestor, the number of descendants per bacterial cell has a heavy-tailed distribution. Surprisingly, selection-like biases emerge, in which common lineages tend to produce more "jackpot" mutations that track lineages with a large number of descendants [5-7]. Independently, Christina Curtis and colleagues have reported 'big bang' expansions of highly successful clonal lineages in tumors despite the absence of selection, again due to a heavy-tailed offspring distribution caused by population growth [8].

These experiments have broader lessons for the genomics of human populations, which are also spatially-structured and have been expanding rapidly for thousands of years [9]. First, population expansion amplifies the effects of stochastic sampling, including mutational jackpot events. Second, such purely stochastic fluctuations can mimic dynamics caused by selection, and have thus been termed "fictitious selection" [6]. When one zooms out from human history, past the proximate causes of events such as the Mongol conquest, the distribution of human Y-chromosome haplotypes appears to be driven largely by chance, or by historically contingent epigenetic factors such as culture and technology, rather than heritable genetic factors.

## Cooperation and kin discrimination has a deep evolutionary basis

Another important parallel with human history and evolution found in evolution experiments is the universal importance of cooperation and kin discrimination. Microbes often produce public goods, such as extracellular enzymes that break down resources such as sucrose [10] or environmental toxins such as hydrogen peroxide [11]. For that reason, microbes have evolved a variety of mechanisms to distinguish kin from unrelated strains. Evolution experiments and studies of the natural history of microbes show that kin discrimination is an ancient social trait, pervasive within and between bacterial populations, species, and communities. Some bacterial strains stab different bacterial strains to death ("contact-dependent killing"), which causes different strains of bacteria to "phase separate", promoting increased cooperation among related strains [12]. Discomfiting parallels with "ethnic cleansing" during many civil wars and genocides are evident.

Bacteria are also capable of sacrificing themselves for the good of their kin. "Suicide bomber" bacteria produce toxins that are costly to produce but disproportionately lethal to unrelated toxin-sensitive strains. By freeing up resources for their kin to use, suicidal altruistic behavior is evolutionarily stable in spatially-structured populations of bacteria, as the benefit to kin outweighs the cost to individual cells [13]. In human cultures, martyrdom and self-sacrifice for the common good is central, and are common elements in religions and political ideologies ("give me liberty or give me death!"). Furthermore, the basis of organized warfare is the commitment of individual soldiers to altruistically sacrifice for their fellow soldiers and the common good—although in practice this often amounts to the spoils of war or access to greater power or wealth in a patronage network.

Experiments with the remarkable social bacterium *Myxococcus* have shown that strains evolved from a common ancestor can rapidly evolve to distinguish each other [14]. In this system, kin discrimination evolves despite the absence of direct selection pressures for that trait, but rather as a side-effect of adaptive evolution to their environment. There is a striking parallel to how human language evolves. Language is crucial not just as a communication medium but also as a marker of cultural belonging and social status. Languages, isolated for long enough, gradually evolve to become mutually unintelligible, and thus can become the basis for kin discrimination, even in the absence of direct selection pressures for mutual unintelligibility.

## Can runaway niche construction lead to ecological suicide?

Another recent surprise is that soil microbes can change the acidity of their environment rapidly enough to drive laboratory populations to extinction, a phenomenon called "ecological suicide" [15]. In another evolution experiment, Michael Travisano and colleagues found increasing rates of population crashes and instability over the course of bacterial adaptation to a stressful environment [16]. Such dynamics may not be constrained to niche construction or ecological change either. In the context of sexual selection, Hanna Kokko has discussed how selection for male reproductive success could result in characteristics harmful to females, and how intense male-male competition could exert costs on females that harm the success of the population as a whole, maladaptively driving a population closer to extinction [17].

These sorts of dynamics are clearly relevant to human populations. Humans are masters of niche construction---altering their own and other species' local environments and thereby transforming natural selection pressures---attested to by agriculture, city construction, and highly technological forms of warfare, all of which could plausibly lead to ecological suicide. Indeed, terms like "anthropogenic climate change" and "nuclear winter" succinctly demonstrate the scale of humankind's influence on the natural environment, and thus their social environments. The ecological changes wrought by industrialization are obviously maladaptive in the long-term, given how trillions of dollars of infrastructure lies on coastlines vulnerable to sea-level rise, and our dependence on ecosystems that are vulnerable to ocean acidification.

Changes in culture, technology, and society are having dramatic effects on the course of evolution on the planet, in particular by accelerating the pace of extinction of natural species, replacing natural habitats with agricultural, industrial, and urban development, and through global changes to the nitrogen and carbon cycle. Will those changes in turn cause moderating feedbacks on human society and behavior? We hope so. Recent work by Daniel Rothman estimates that the amount of carbon required to trigger a catastrophic mass extinction is roughly that projected to be produced by humanity around the year 2100 [18]. An open question is whether global civilization—itself a complex adaptive system—will be able to adapt on a short enough timescale to avoid catastrophe [19].

## Calibrating our intuitions by comparison to evolution experiments

An interesting premise in teaching evolution to students is to compare timescales for evolutionary change in human populations to the timescales of evolutionary change in laboratory experiments. We can come up with a rough comparison with reference to Richard Lenski's long-term evolution experiment [20]. Lenski's experiment involves 12 populations of *Escherichia coli*, founded from a common ancestor 30 years ago, that each have been evolving in isolation for 70,000 bacterial generations. Each day, 1/100 of each population is diluted into fresh growth media. So, each population goes through $\log_2(100)$ doubling each day, or ~6.6 bacterial generations. Each bacterial population roughly has an effective population size of 5,000,000 bacteria, which is the bottleneck that occurs during the transfer into fresh growth media. If we assume that one human generation takes 30 years, then each day of Lenski's experiment represents about 200 years of human evolution.

In each replicate population, beneficial mutations occur, increase in frequency, and compete for eventual establishment. A natural timescale for evolution in this experiment is the time that it takes for a beneficial mutation to occur and establish in the full population. This is known as a "fixation" event, because then all future individuals in the population have that beneficial mutation fixed in their genomes.

This is also called a "selective sweep", because the beneficial mutation sweeps out all genetic variation found in competing strains.

In one population of this long-term experiment, the time for the first selective sweep was 2,500 generations, although large changes in gene frequency occur within 500 generations. Translating into human generations, this represents 75,000 years for a selective sweep (2,500 generations), and 15,000 years for an equivalent large change in gene frequency (500 generations).

One caveat with this comparison is that in clonal populations of bacteria, selective sweep can be prolonged by competition between beneficial mutants [21-23]. In a separate experiment with budding yeast in which sexual recombination was an experimental treatment [24], an early selective sweep took about 300 generations to complete. Again assuming 30 years for one human generation, this represent 9,000 years. We emphasize that these estimates are crude at best: in an evolution experiment with sexual recombination and adaptation from pre-existing genetic variation, Kosheleva and Desai found that population fitness increases in the absence of complete selective sweeps [25].

Are these crude estimates at all relevant to human populations? As a rough check, we can compare these numbers to current estimates of the time passed since the most recent common ancestor for mitochondrial DNA (mitochondrial Eve) and all Y-chromosome haplogroups (Y-chromosomal Adam). Estimates for the time passed since mitochondrial Eve is 99,000-148,000 years ago or 3300-4940 generations, and 120,000-156,000 years or 4000-5200 generations for the time passed since Y-chromosomal Adam [26]. These estimates are consistent with the recent African origin of modern humans before a series of migrations out of Africa from ~50,000 to ~135,000 years ago [27].

In contrast, linguistic and cultural change occurs far more rapidly. As a historical comparison, 10 generations is 2 days for bacteria in Lenski's experiment, and 300 years for humans. In 1718, the world population was ~600,000,000 people. 100 generations is 15 days in Lenski's experiment, and 3000 years for humans. 1000 BCE is around the time of the Late Bronze Age collapse, and the world population was ~50,000,000 people. 1000 generations is 151 days of evolution in Lenski's experiment (5 months), and 30,000 years for humans. 28,000 BCE was the Upper Paleolithic, and world population was ~2,000,000 people. In short, for cultural and linguistic change to effect genetic change through selection, social norms have to be sustained over an extremely long time. Currently, the few social norms that seem to have been sustained long-enough to provoke genetic evolutionary change through selection are those that aided survival during the thousands of generations that humans lived as bands of hunter-gatherers and, more recently, those related to the settlement of humans into either herding or sedentary lifestyles after the invention of agriculture and animal husbandry during the Neolithic revolution ~10,000 years ago [28].

### *Comparison of timescales for demographic change and admixture*

In contrast to the timescale for selective sweeps, evolutionary change due to recombination, gene flow, migration, and admixture between populations occurs on faster timescales. Even though these processes dominate bacterial evolution in nature [29, 30], so far few experimentalists have studied them in the lab [24, 25, 31-33]. The dynamics of migration and admixture are central to human history, evidenced by population genomics and breakthroughs in the sequencing of ancient DNA from archaeological and fossil remains [4]. For instance, significant admixture has occurred in the Americas in the past 500 years, due to waves of voluntary and forced migration [34]. The arrival of Europeans in the Americas resulted in the rapid decline of Native American populations, due to the importation of Old World plagues such as smallpox and measles [35]. In much of the Americas, natives were enslaved by Europeans and compelled to work on plantations. When those natives died due to overwork, poor nutrition, and lack of resistance to Old World diseases, slaves from Africa were imported as replacement labor, leading to sometimes novel, sometimes recurring admixture among various Native American, European and African populations. Strict racial hierarchies were put into place in Spanish, Portuguese, French and British colonies based on different degrees of European descent (i.e. [36]). This history has left a strong imprint on contemporary culture in the Americas: contemporary social and political order, as in the past, is largely based on race and degrees of European descent [37]. This fact is the basis of many contemporary political pathologies in the United States. For instance, it is possible that one reason why funding for public education and healthcare has been cut over and over again in the past decades, is spiteful political motivations driven by anxieties over race. Later in this paper, we discuss such issues in the context of cultural and institutional evolution. Feedbacks between demographic change and cultural and institutional evolution remain an important topic for future research that is highly relevant to contemporary politics and policy.

## Non-genetic modes of inheritance in humans obey evolutionary dynamics

There are obvious limitations to mapping the dynamics of human history and evolution onto evolution experiments. Genetic evolution is slow, and while social dilemmas and public goods games have been studied extensively in evolution experiments [38-40], genetic evolution has little to do with everyday social dilemmas in which community welfare is at odds with individual incentives [41], or in social dilemmas in which particular groups profit from the exclusion or oppression of other groups at the expense of overall welfare. Nevertheless, the social tools that humans have to deal with social dilemmas such as racism and climate change themselves are subject to evolutionary dynamics that operate on much faster timescales: language, behavior, culture, and symbolic communication [42]. This occurs because languages, behaviors, cultures, and symbolic communication are transmitted across human generations (they are heritable), they vary within and across populations (phenotypic variation), and different variations may spread faster, be copied or imitated more, or be more resilient against the invasion of other variations (selection). For this reason, there is significant conceptual overlap between historical issues that come up in the social sciences and in evolutionary biology and computation. We discuss some of these conceptual overlaps next.

## Conceptual overlaps between evolutionary experiments and computation with social science

### *Gene-culture coevolution is important in human history and evolution*

That we have culture and symbolic communication is in part dependent on a series of genetically evolved characteristics [43]. In turn, culture can alter the course of genetic evolution through niche construction [44]. The challenges of everyday life in the past has honed a species that is both capable of aggressively seeking individual gain as well as banding together to solve collective action problems. Ancestors with a greater ability to coordinate and therefore address collective problems had an evolutionary advantage over those who did not [45, 46]. Coordination has biological prerequisites, such as the ability to estimate the intentions of other actors, and remember an approximate accounting of others' past behavior, both collaborative [47] and antagonistic [48]. Facial recognition and emotional interpretation, for example, are of tremendous help in developing and navigating complex societies. In turn, cultural innovation has over hundreds of thousands of years played a role in genetic innovation. For instance, hominids such as *Homo habilus* and *Homo erectus* were able to make sophisticated stone tools that probably helped them survive in their environment. To give another example, the domestication of cattle, goats, and camels for milk in parallel across cultures imposed strong selection pressures for the ability to digest lactose. In this manner, genes and culture co-evolve [49].

Similarly, does epigenetic information play any role in evolution experiments and computation? Although once controversial, it seems so. For example, nongenetic variation in the ability of bacteriophage λ to bind alternative host receptors was important in catalyzing an evolutionary innovation in which the phage protein evolved a qualitatively new function [50]. Interestingly, evolutionary algorithms inspired by epigenetic inheritance have found practical use in solving multi-objective optimization problems [51, 52].

## *Adaptation, chance, and historical contingency in culture and human institutions*

A fundamental question in evolutionary biology is the relative roles of adaptation, chance, and historical contingency (i.e. historical accident) in shaping ensembles of possible evolutionary outcomes and histories. Although this question cannot be answered for singular evolutionary trajectories, such as the origin and subsequent evolution of life on Earth, evolution experiments allow replicate evolutionary histories to be run and replayed, such that the contributions of adaptation, chance, and contingency in evolution can be teased apart [53, 54]. Adaptation, chance, and historical contingency are important factors in any historical process, and for that reason are highly relevant in the study of cultural and institutional change and design.

Institutions help define communities by organizing and transmitting information that affects individual phenotype and behavior through learning, imitation, and inculcation [43, 44]. Often, institutions turn on a *deontic* that allows, obliges, or forbids some behavior. Transgressors can expect, with some probability, a cost (the formal stipulation of which transforms a norm into a rule). These costs can be externally and internally imposed, and the former begets the latter, as when—after years of scolding from your mother about, say, biting your fingernails—you feel guilty when engaging in the "offensive" behavior. In some cases, such institutions are good for both society and in the individual, as in prohibitions against drinking alcohol, smoking cigarettes, or using drugs. And in generating a regularity of human behavior, institutions ease communal interactions, freeing individual cognitive bandwidth for other tasks. There are plenty of instances, however, where a deontic unduly burdens some members of society but not others, as either chauvinism or self-hatred in the form of overt and eventually internalized misogyny, anti-Semitism, or racism.

Inculcation can be seen as matter of operant conditioning involving rewards and punishment. That is, norms and rules may be transmitted through the promise of boosted payoffs, rewards, and warm fuzzy feelings brought on by community approval and the corresponding sense of virtuousness, or by the fear of punishment or sanctions. In the latter case, the teaching and inculcation process may not necessarily be for the learner's benefit, but for the benefit of the teachers, and perhaps (hopefully) the larger community. One's preferred strategy for a given game, for example, may require cooperation from other players and so one and one's fellow adherents have a stake in perpetuating one's strategy through sanctioning and inculcation. Whether or not the spread of this strategy leads to an increase in payoffs for converts, or to an increase in social welfare, depends on the game.

But where do institutions come from in the first place? Humans routinely encounter strategic, multi-party situations that are liable to repeat, and—as game theoretic folk theorems have shown—such repeated scenarios have myriad viable, stable equilibria. That one equilibrium is settled on over others in a given community may occur for any number of reasons. A particular strategy might be chosen because of what it means for linked or nested games, or because of path dependencies and historical trajectories (e.g., [55]), for matters of convenience, or for no particular reason whatsoever. Over time patterns become set, a regularity of behavior is established among participants, and the arrangements become a tacit part of the community, enabling individuals to (near-unthinkingly) carry out tasks without endlessly renegotiating what might otherwise be thorny communal issues [56].

By invoking a deontic—a "may", "must", or "must not"—norms and rules blur the line between prudent and proper action. Some institutions are blatantly moral, dealing with who gets what, and why. The crucial concept of desert—what it means to deserve some resource, treatment, or responsibility—is itself an emergent social institution [57] that can be meaningfully translated into and from the moral grammar of justice, and has a direct effect on the distribution of power and wealth in a society [58]. But by specifying acceptable, unacceptable, and obliged behaviors whose observance or violation in turn requires others to monitor, punish, and reward, even ostensibly pragmatic norms and rules take on a distinctly moral flavor. In their weakest formulation, such institutions deal with manners and niceties. Stronger formulations amount to what we often call duties.

All people have behaviors they consider crude, questionable, and taboo, polite, noble, and obligatory, but where any specific behavior falls among these categories changes across communities, sometimes drastically (e.g., [59, 60]). Deontics can cause institutions that arise in response to particular environmental or social challenge to become entrenched even in the absence of the original reason for that institution, such as the ceremonial monarchies that exist in many countries. Deontics thus create a positive feedback loop that maintains a "memory" of historical conventions. Historical "memory" affects all evolving systems, since heritability is defined as phenotypic correlations between past and present generations. How quickly does the imprint of the past decay in evolving systems? In what circumstances is "memory" of the past adaptive, say, as pre-adaptations to recurring conditions? And in what circumstances is "memory" of the past maladaptive, say, when entrenched beliefs or adaptations to historical conditions inhibit adaptation to present or future conditions?

## *Entrenchment and contingency in cultural evolution*

As a shortcut to the process of learning from direct experience in each generation, inherited institutions allow humans to learn about the world from others around them. This shortcut, however, comes at a price, namely: individuals who inherit the institution may not fully grasp its relationship to the problem it helped solve. Although an obvious advantage overall, this may lead to all manner of trouble, including the aggrandizement of the institution to sacred rather than sensible causes, and the misapplication of the institution to unrelated problems. We sketch out a hypothetical example to illustrate our point.

### Deontics stabilize multiple equilibria in indefinitely repeated social dilemmas

Imagine two rivers with people living along the length of each. Upstream families can divert the river such that downstream families are forever worried about scarcity of the common-pool resource. Along River A, people adopt an obsequious demeanor when interacting with upstream families, devoting resources to currying their favor by delivering gifts, while the people of River B strike an aggressive posture and spend resources to maintain a credible threat of violence. Because everyone except the distant-most river residents has an upstream and downstream neighbor, both arrangements are in equilibrium, both keep the water flowing, and both require continued investment from all parties.[1] The overall ethoses, however, are very different.

The founding people of River A are in not in any fundamental way different—friendlier or more docile—than the founding people of River B. Early generations devised the now dominant strategies through feedback and learning, perhaps trying many solutions before discovering one that stuck. The people of River A might have become as warlike as those along River B had someone successfully experimented with warfare before testing tribute and trade. People can see their neighbor's water level and correlate it with that neighbor's strategy. For an array of reasons, someone along River A successfully tried gift-giving before fighting, and the behavior spread, saving neighbors the costs of further experimentation, which may be substantial when it comes to water that one's family needs now. Early on, the respective peoples of River A and B *adopted* friendly or antagonistic attitudes in their political relations. Later on, the peoples of River A and B *were raised* to be friendly or antagonistic, at least within the context of upstream relations.

---

[1] This toy example scales up easily enough, so that the cooperative residents along River A work to jointly manage water flow, and the truculent residents of River B settle into a pattern of coalition-building and warfare; a sort of balance-of-powers, whereby various downstream neighbors ally and oppose various upstream groups through threat of force.

The crucial idea is that the first peoples used feedback from the common resource—water level in this case—to determine whether or not the behavior boosted payoffs. This is not necessarily so for subsequent generations, who become indoctrinated into the initial behavior. Early generations of the people alongside Rivers A and B learned that upstream gift-giving or threat-making led to a stable water flow. Following generations may have noticed the association between water level and gift-giving or threat-making, but what they definitely learned was that failure to observe gift-giving or threat-making norms resulted in dirty looks and other sanctions from their family members. Entering as they are into an equilibrium with the payoff of a reliable water flow, they may never directly observe the strategy's relationship with the resource. Early generations knew they were amiable or hostile toward upstream people *to keep the water flowing*. Following generations know that they are amiable or hostile *to avoid reprimand* from others in their community. For both generational classes, the actors are looking to avoid a decrease in payoffs. In the former, that decrease primarily comes from competition over the resource. In the latter, the decrease may partly come from the resource, but foremost in peoples' minds are decreases through communal-sanctioning and self-sanctioning. Institutional economists often refer to these cost as external and internal delta parameters, respectively.

What was at first a narrow, prudent aim has been replaced by a broad, moral imperative. Niceness along River A and nastiness along River B are over time less valued for their effectiveness at maintaining water levels and are increasingly associated with the community's identity and character, valued in and of itself as subsequent generations perpetuate the norm that they were raised to. Such are the benefits and costs of inheriting an equilibrium rather than forging it: one learns a lesson, but not the original incentive for it. You know you are supposed to be nice or mean to the neighbors upstream, without the knowledge of why. You never went without water for lack of being nice or mean. You just got yelled at by your parents. You inherit the strategy, but not the thirst that originally motivated it.

As in our example, cross-cultural psychologists find that the ethos of a community is deeply tied to the biophysical environment and natural resources that sustain it. For example, cross-cultural psychologists have found evidence that people from pastoralist lineages are more prone to violence that are people descended from farming communities [61] and cultures dependent on rice cultivation appear to be more interdependent than those that grow wheat [62]. A key difference between these cases and the hypothetical example we give is that one of a multiplicity of equally plausible solutions can become culturally entrenched over time, without the need for greater collective cooperation in one over the other, as is required to grow rice compared to wheat. On top of showing that a common biophysical environment can lead to widely divergent dispositions toward cooperation between populations, our hypothetical scenario further illustrates *how* a seemingly pragmatic, non-normative institution can grow in scope and cultural importance through the untethering of deontic from the problems their institutions evolved to address.

## Exaptation in cultural evolution

Once its initial aim become obscure over time, we suspect that inherited institutions will often outlive the problem they evolved to address. In our upstream/downstream example, it is easy to see how deference and aggressiveness may change from behavioral strategies into communally-valued traits. Institutions addressing particularly salient problems like water supply are wont to be especially dogged, with early generations firmly implanting the behavior, and the tools for its future observance, in their children. It is no great leap, then, to see that beliefs strongly reinforced by the community yet mentally untethered from their *raison d'etre* might be repurposed and applied to any number of problems. Or, if not directly applied to new problems, they may be thought of as appropriate behavior in new scenarios to no benefit beyond the aforementioned role of all institutions in generating social regularity and clear interpersonal expectation, and possibly even to individual and communal detriment.

## Secondary contact in cultural evolution

In addition to being used in novel scenarios, inherited institutions may be applied in the same setting but with a twist: contact with another group that has evolved a different solution to the problem in question. If River A and River B turn out to be two distant sections on a very long river, the peoples from each may eventually encounter one another. As observers of this hypothetical clash of cultures, we may root for the success of one over another, but from the narrow perspective focused on continued water flow, the spread of either the fawning (pacific) or assertive (aggressive) culture will suffice. Rather than a rational comparison of the two equilibrium-inducing solutions to the downstream problem, though, conflict may arise as a matter of *identity*, with River A and River B people vying not just for the spread of their downstream solution but for the spread of what that solution has become: their morals and communal character. The story of their encounter, however, may not be a violent clash. The fundamental alteration to the social environment, once incorporated into the evolutionary game, means that the process may yield a whole new equilibrium wherein both cultures are transformed.

Migration and emigration between populations surely plays an important role in cultural evolution, as it does in genetic evolution. Understanding how cultural differences causes permeable barriers to admixture, say due to taboos in marriage across class, race, caste, or community lines, and how genetic admixture or (lack thereof) plays a role in cultural evolution is poorly understood and ripe for study.

## Future directions and possible contacts with the social sciences

Evolution experiments can only have contact with the social sciences insofar as data from both fields fit the same classes of mathematical models. This is occasionally possible because both fields share a common toolkit in evolutionary game theory and dynamical systems, as well as a common interest in the evolution of cooperation and pro-social behavior [63-71]. Despite the attention given to prisoner's dilemmas, scholars in the social sciences have taken seriously Long's (1958) [72] observation that humans—and all social animals—participate in a diversity of games, studying reciprocity, cooperation, and trust in a variety of scenarios, including hawk-dove [73], stag hunt [74, 75], public goods [76], and common-pool resource settings [77]. Similar questions have also been asked in the context of both the evolutionary biology of social behavior and more narrowly, in the context of evolutionary experiments with microbes and computer algorithms [38, 78, 79]. There are also other key points of overlap between the evolutionary and social sciences. In both populations and social organizations, the distribution of phenotypes, behaviors and strategies are shaped by the environment (natural selection or sexual selection in populations versus conformity and economic payoffs in social organizations) but a diversity of non-harmful or less-than-optimal phenotypes often persists. In both populations and in social organizations, there is often a delay between environmental change and adaptation, causing an evolutionary mismatch [80] such that phenotypes, behaviors and strategies outlive their environment.

The key difference between the social sciences and the study of genetic change in evolution, is the importance of transmissible epigenetic information and niche construction in the evolutionary dynamics of human culture [49, 81]. Culture is contingent on evolved traits that allow the capacity for language and intergenerational memory [43] and in turn culture can directly change short-term evolutionary dynamics by affecting probabilities of survival and reproduction [44, 49]. Hence, for evolution experiments and computation to have sustained relevance to the social sciences, model systems with dynamics that depend on both epigenetic inheritance as well as genetic inheritance, niche construction, and feedbacks between epigenetic inheritance and genetic inheritance (for example, mate choice) are needed. What questions may be worth asking? We sketch some possible directions below.

### *How do institutions, technological and ethical norms, and social conventions evolve or change over time?*

A fascinating open question is how ethical norms perpetuate and change over time, and how they might affect how social organizations adapt to fluctuating or changing environmental conditions. Ehrlich and Levin [82], correctly in our opinion, propose that studying the dynamic interplay between individual behavior and normative rules may be key in understanding sudden phase transitions that can change the cultural landscape. In the face of ongoing climate change and the lack of political will to confront the ecological crisis, few questions are more important for civilization to survive the next 1000 years.

Such phase transitions in social and ecological systems often manifest as tipping points. Recently, Damon Centola and colleagues [83] reported evidence of a tipping point in an experimental system studying social coordination. In this work, a critical mass of a committed minority of participants (~25%) was required to flip the state of an arbitrary naming convention.

Ehrlich and Levin [82] have hypothesized that predominantly vertically transmitted norms, like ethical values, are "sticky" or "highly viscous" in how they flow over time in an evolving population. How might such viscosity affect tipping points in social systems? We suggest that sticky norms might cause delays in system change in response to rapid environmental change. That is because ethical beliefs imply behavioral obligations that make them self-enforcing: such norms perpetuate through external costs imposed by individuals on others who violate the norm (i.e., sanctioning) and through internal costs that individuals who violate the norms impose on themselves (i.e., inculcation, which leads to guilt and is likely built on the expectation of sanctioning). On the other hand, sticky norms might cause institutions, cultures, and societies to remain robust in the face of temporary threats from rivals or temporary environmental disturbances such as natural disasters. More intriguingly, Simon Levin [84] has proposed "mutual coercion, mutually agreed upon" as a workable solution to avert tragedies of the commons, such as climate change. To that end, better ethical norms may be an essential part of solving public goods problems at scale.

## *Why are institutions, laws, and morals often asymmetrical in their application?*

It is commonly taken for granted today that laws and codified rules should be applied without regard to race, gender, sexual orientation, class, or ability. Not all rules are formal, however, and plenty of rules on the books are not rules in practice. Regarding those that are, there are nonetheless strong asymmetries in their application. This observation, that many institutions actively propagate white supremacy or the oppression of women is the basis for critical race theory [85] and feminist legal theory [86]. Examples of asymmetrically applied moral principles include "drug addicts deserve to be treated as criminals," which was a common belief during the heroin and crack epidemics in the 1980s that predominantly affected black populations in the U.S., but is starting to shift now that the ongoing synthetic opioid epidemic is predominantly killing white populations in the U.S.

Ironically, one of the causes of the synthetic opioid epidemic is the symmetric application of the position that "selling addictive drugs to satisfy foreign demand is acceptable" by Chinese pharmaceutical companies, and by the market system more broadly. Chinese modern history begins with the Opium Wars, in which the British waged war against China in order to sell opium to the Chinese to maintain a favorable balance of trade. Certainly, Chinese companies selling Fentanyl to the U.S. are aware of both the consequences of their actions as well as the historical irony in applying a moral rule symmetrically.

On the basis of first principles, the inconsistent application of moral and legal principles is a contradiction. But from the perspective of institutions and morals arising from a process of cultural evolution, asymmetry emerges naturally, especially from the perspective of kin discrimination. Such a perspective, however, is descriptive (i.e. what happens in the world) and not prescriptive (the way things ought to be) since an evolutionary perspective on institutions and legal principles would necessarily concern itself with their consequences (i.e. norms and beliefs change in response to their outcome) rather than with justice as an immutable, universally defined concept. Put differently, students of evolution have much to say about morals if they approach desert and related concepts, including justice, as emergent social institutions rather than inborn predispositions or metaphysical constructs, as dominant psychological and philosophical camps contend [57, 58, 87] Rather than something that simply exists, justice can be meaningfully approached as something we create.

## *How does ecological inheritance affect evolutionary and social outcomes?*

It is a truism that the ability of a population of organisms to survive and reproduce depends on their ecological context. Consider a population of populations, in which some subpopulations occur in environments with a higher carrying capacity than others. Naively, it would seem that given proportionate migration, organisms in better environments (say with higher carrying capacity) would have higher fitness—even in the absence of any genetic determinants—than those in worse environments. Ecological conditions can also be heritable, such as for organisms that hold territory and engage in niche construction. To our knowledge these questions have been little studied in the context of evolution experiments with microbes and on the computer.

Certainly, ecological inheritance must play some important role in both cultural evolution and niche construction in humans as well as diverse organisms in the natural world. For instance, differences in economic capital between individuals tends to amplify over time due to the exponential growth of capital. These issues are highly relevant in contemporary society. In a widely reported study, Raj Chetty and colleagues reported that conditioning on parental income, wealth and privilege is far more heritable for white men than black men in the United States. This result is driven by differences in wages and employment status; notably these differences do not occur between white and black women [88]. Widening economic inequality in the United States is a serious social issue: rates of social mobility in the United States has dropped from ~90% for children born in 1940 to ~50% for children born in the 1980s. This result is robust to many economic assumptions; furthermore, increasing GDP growth rates ("growth-friendly policy" or "trickle-down economics") cannot restore mobility rates to that experienced in the 1940s. In contrast, changing the distribution of growth across income groups to the more equal distribution experienced by the 1940 birth cohort would reverse more than 70% of the decline in social mobility [89]. Moreover, rising economic inequality fosters the subversion of legal, political, and regulatory institutions by the wealthy and politically powerful for their own benefit [90].

## *Positive frequency-dependent selection and multiple equilibria in social systems*

A key dynamic affecting the adoption of social norms is the presence of multiple steady states or equilibria maintained by positive frequency-dependent selection. For instance, in the United States people drive on the right-hand side of the road, while people in the United Kingdom drive on the left-hand side of the road. Following the dominant norm in each country is strongly advised. Switching is possible, but only with en-masse coordination, as was done by Sweden on Högertrafikomläggningen ("The right-hand traffic diversion") in 1967. As another example, switching from an economy based on fossil fuels to one based on carbon-neutral energy sources is necessary but difficult due to the costs of social coordination.

Positive frequency-dependent selection, caused by a need for social coordination, also occurs in microbial systems. For instance, Olaya Rendueles and colleagues [91] found pervasive positive frequency-dependent selection among isolates from a centimeter-scale population of the social bacterium *Myxococcus xanthus*. Strains that were poor competitors at intermediate frequencies were competitively dominant at high frequencies. Rendueles *et al.* further demonstrated that positive-frequency-dependent selection maintains diversity in patchily distributed populations, suggesting that positive frequency dependence contributes to *Myxococcus* diversity by reinforcing social barriers to cross-territory invasion and promoting within-group relatedness. Surely, such dynamics need to be investigated more fully in evolution experiments with microbes and on the computer, given its relevance in both natural populations as well as for social behavior in general.

## *Do key innovations ever play a role in evolutionary rescue?*

In the absence of any political will to adequately respond to ongoing climate change, one hopes for some sort of “Hail Mary” gambit, such as an improbable technical innovation like nuclear fusion that makes a fossil-fuels-based economy obsolete without political or economic sacrifice.

Evolution experiments have serendipitously become model systems for the study of key innovations in evolution [92, 93]. Can key innovations destabilize equilibria caused by positive-frequency dependent selection? Do key innovations ever rescue populations from extinction, say caused by ecological change or environmental degradation? Perhaps these questions could be answered with the right microbial system, or on the computer.

## *Stochasticity in cultural evolution for rapidly expanding populations*

As we have discussed earlier, recent research shows that stochastic events have unexpectedly large effects on the evolutionary fates of expanding populations. Might the same be true for cultural evolution during the adoption of alternative social norms? We hypothesize that under conditions of expanding populations, stochastic choices in which norms are chosen early in time can have large evolutionary outcomes (at least on a timescale of tens of generations), because initial conditions due to random chance can be amplified. In analogy to the results found in experiments with bacterial colonies and tracking tumor growth in model systems, we predict that this is a purely 'neutral' outcome that from a dynamic point of view may masquerade as selection. Perhaps a real-world example might be the adoption of western dress (i.e. business suits) across much of the global business world.

## *To what extent does epigenetic and ecological inheritance account for trait heritability?*

An important scientific problem today, is that estimates of the heritability of traits (i.e. regression of a trait between offspring and parents) could be confounded by the effects of heritable but non-genetic factors, such as local environment (such as air or water pollution), epigenetic effects such as maternal effects [94], wealth (including access to quality healthcare or education), and culture. For example, Feldman and Ramachandran [95] show that incorporating cultural inheritance into models for the determination of phenotypes sharply reduces estimates of the genetic contribution to heritable phenotypes. Given that researchers have built, and will continue to build, models of the contribution of genetics to intelligence, children's educational achievement, and economic and political preferences, neglecting cultural inheritance and heritable differences in capital and social status would clearly cause serious problems. Importantly, the magnitude of cultural effects can be comparable to the magnitude of genetic effects [95].

Models of trait heritability including epigenetic sources of inheritance stand as a potential, non-mutually exclusive alternative to the recently proposed 'omnigenic' model for the genetic basis of complex traits [96] in order to explain why genome-wide association studies have largely failed to explain the basis for human disease susceptibility.

## *What role does spite play in climate change, niche construction, economic markets, and public goods games?*

In the context of evolutionary game theory, spite is defined as a suffering a cost in order to exact a larger cost on someone else. For example, a population of organisms that produces an environmental toxin at some cost can still benefit if the toxin kills off sensitive competitors [13]. It is possible that to some degree, resistance to action to mitigate climate change might be spiteful. That is, if the consequences of climate change are largely foisted on other nations or geopolitical enemies, then the costs facing some nations could be offset by the benefits of hurting competing nations and interests. Notions of ecological justice [97], are highly relevant here, since often particular cultures, ethnic or racial groups suffer disproportionately from externalities such as pollution produced by competing interests.

Spite is certainly relevant to public goods games [84] but more work needs to be done to understand the degree to which it affects the management of the global commons, such as the oceans and atmosphere, which are already vulnerable to tragedies of the commons. We propose that spite in fact plays an important role, given how the U.S., a major carbon polluter, has withdrawn from the Paris Climate Accords as part of its 'America First' policy despite global opprobrium.

## Conclusion

Evolution experiments with microbes and computer programs have proven to be a powerful paradigm for studying general evolutionary processes and dynamics important in ecology, evolutionary biology, computer science, and engineering. Here we argue that evolution experiments also shed light on processes traditionally studied by the social sciences. Given the substantial overlap in both mathematical and conceptual tools across fields, we believe that empirical and theoretical progress in each can inform and inspire progress in the other. A key difference between fields is the importance of evolutionary processes that involve multiple layers of inheritance (such as genes, epigenetic inheritance, language, and culture) and feedbacks (such as gene-culture coevolution and mate choice) as well as open-ended niche construction in the social sciences. Many of these topics are highly active research areas in evolutionary biology and the social sciences. In particular, we believe that computational models of niche construction may allow for the open-ended evolution that continues to be a goal in computational evolution experiments. Furthermore, biological and computational experiments that are tied to phenomena in the social sciences may be an exciting way to motivate evolution and computation to scholars and students across diverse cultural and socioeconomic backgrounds, as well as to scholars and students in the social sciences and humanities. Given mass surveillance of citizens over social media and the internet by corporations and governments, computational models of cultural evolution in principle could be compared to data, given that large-scale datasets already exist for tracking cultural change in real-time. Thus, experimental evolution, as a laboratory and computational science, is poised to grow as an educational tool for people to question and study where we come from, why we believe what we believe, and where we as a species may be headed.